\documentclass[aps,preprint,nofootinbib]{revtex4}%
\usepackage{amsfonts}
\usepackage{amsmath}
\usepackage{amssymb}
\usepackage{graphicx}%
\providecommand{\U}[1]{\protect\rule{.1in}{.1in}}

\begin{document}
\preprint{ }
\title[Short title for running header]{Comment on \textquotedblleft Gauge Symmetries and Dirac
Conjecture\textquotedblright\ by Y.-L. Wang, Z.-P. Li, K. Wang and some other
counterexamples to the Dirac conjecture}
\author{N. Kiriushcheva}
\email{nkiriush@uwo.ca}
\author{P. G. Komorowski}
\email{pkomoro@uwo.ca}
\author{S. V. Kuzmin}
\email{skuzmin@uwo.ca}
\affiliation{The Department of Applied Mathematics, The University of Western Ontario,
London, Ontario, N6A 5B7, Canada}

\begin{abstract}
We argue that the conclusion about the invalidity of the Dirac conjecture,
made in the paper by Wang, Li, and Wang (\textit{Int. J. Theor. Phys.
}\textbf{48 }1894, 2009), was based upon a flawed analysis of the proposed
counterexamples. In the case of the Maxwell theory, the well-known gauge
symmetry is contradicted by the results in the Hamiltonian and Lagrangian
approaches presented by the authors. We also consider the oldest
counterexample to the Dirac conjecture due to Allcock and present its natural parametrization.

\end{abstract}
\volumeyear{year}
\volumenumber{number}
\issuenumber{number}
\eid{identifier}
\date{\today}
\received{}

\maketitle

\section{Introduction}

From the time of the formulation of Dirac's conjecture (1964) \cite{Diracbook}
the question of its validity has sporadically appeared in the literature. One
recent example is the paper of Wang, Li, and Wang \cite{WangLiWang2009} in
which they argued against the validity of Dirac's conjecture on the basis of a
few examples. A re-examination of these examples is the main subject of our paper.

The conjecture made by Dirac concerns the connection between gauge symmetries
and the first-class constraints that may arise in the Hamiltonian formulation
of singular Lagrangians (not all singular Lagrangians are gauge invariant),
i.e. all first-class constraints generate gauge transformations, or the
knowledge of the full set of first-class constraints is sufficient to restore
the gauge transformations.

In 1982 Castellani \cite{Castellani} converted Dirac's conjecture into a
theorem (see the bottom of page 364) that is based on a detailed description
of the procedure to build a gauge generator using all of the first-class
constraints (second-class constraints must be eliminated first). In
\cite{Castellani} this procedure was originally illustrated in the Hamiltonian
formulation by using two examples: Yang-Mills (in its original variables) and
Einstein-Hilbert (in the ADM parametrisation). Castellani excluded the case of
$\chi^{n}$-type first-class constraints (powers of constraints) for which the
best known example is due to Cawley \cite{Cawley1979}; we have deferred its
discussion to future publications because the examples considered in
\cite{WangLiWang2009} are not of this type.

The availability of the Castellani procedure has made the question of the
correctness of Dirac's conjecture a pure technical exercise (although
potentially cumbersome): one must find all first-class constraints
(second-class, if any, must be eliminated); build a generator; find the
transformations of the phase-space variables from which transformations in
configurational space can be derived\footnote{This is not always the case, but
the models considered in this paper do not have this sort of complication.};
and check if the resulting gauge symmetry is also a symmetry of the initial
Lagrangian. If by using all first-class constraints the gauge transformation
is found, and it is also a gauge transformation of the original Lagrangian,
then the conjecture is valid. This approach was attempted in
\cite{WangLiWang2009}, but the calculations are incorrect; therefore, the
conclusions based on such calculations must be wrong.

Whether or not a particular gauge transformation is a symmetry of the
Lagrangian can be checked directly; although it can be a difficult task for
complicated models. Fortunately, one can rely on Noether's identities (also
known as Differential Identities (DIs)), which are linear combinations of
Euler-Lagrange Derivatives (ELDs) that are identically zero. This is not an
on-shell\ condition, and the solution of the equations of motion ($ELD=0$) is
irrelevant. This off-shell condition (DI) in the Lagrangian approach is
compatible with the following view on Dirac's conjecture expressed in
\cite{Frenkel1980}: \textquotedblleft The value of Dirac's conjecture would
have been that it would give all the gauge generators without dealing
explicitly with the often complicated equations of motion\textquotedblright.

The paper is organised as follows. In the next Section we reconsider all
(three) examples discussed in \cite{WangLiWang2009}, in detail. In Section 3
we study different parametrisations of the oldest counterexample to the Dirac
conjecture, which is due to Allcock \cite{Allcock1975}, and show how the
simplest, natural parametrisation can be found. In Conclusion we summarise the
results of this paper and briefly outline further directions and expectations.

\section{Three examples -- three unjustified conclusions}

The authors of \cite{WangLiWang2009} claimed that two new counterexamples to
the Dirac conjecture have been devised, while using the well-known Maxwell
Lagrangian they illustrate a case in which the Dirac conjecture is valid. We
shall consider and analyse each of the examples.

\subsection{First example}

The first counterexample considered in \cite{WangLiWang2009} (also see the
references therein to the earlier work of the authors) is a model with the
following Lagrangian (Eq. (WLW16) of \cite{WangLiWang2009})\footnote{In this
Section, equations from \cite{WangLiWang2009} are indicated as Eq.
(WLW\#\#).}
\begin{equation}
L\left(  x,y,z\right)  =\frac{1}{2}\left(  e^{2u\left(  y\right)  }\dot{x}%
^{2}+e^{-2v\left(  -y\right)  }\dot{z}^{2}\right)  ,\label{eqn4}%
\end{equation}
where a dot indicates the time derivative, and $u\left(  y\right)  $ and
$v\left(  -y\right)  $ satisfy the following equations (Eqs. (WLW17) and
(WLW18))\footnote{Note that ordinary differential equations (\ref{eqn5}%
)-(\ref{eqn6}) have simple solutions, but we keep this form to be close to the
presentation in \cite{WangLiWang2009}.}%
\begin{equation}
u^{\prime\prime}\left(  y\right)  =u^{\prime}\left(  y\right)  +2\left[
u^{\prime}\left(  y\right)  \right]  ^{2},\label{eqn5}%
\end{equation}%
\begin{equation}
-v^{\prime\prime}\left(  -y\right)  =v^{\prime}\left(  -y\right)  +2\left[
v^{\prime}\left(  -y\right)  \right]  ^{2}\label{eqn6}%
\end{equation}
(where $u^{\prime}\left(  y\right)  =\frac{du\left(  y\right)  }{dy}$ and
$v^{\prime}\left(  -y\right)  =\frac{dv\left(  -y\right)  }{d\left(
-y\right)  }$). In \cite{WangLiWang2009} it is claimed that the gauge
transformations obtained by the Dirac approach, when it is applied to
(\ref{eqn4}), differ from the Lagrangian symmetry. To check this claim, let us
first perform a Hamiltonian analysis of this Lagrangian.

To obtain the total Hamiltonian, we perform the Legendre transformation%
\begin{equation}
H_{T}=p_{x}\dot{x}+p_{y}\dot{y}+p_{z}\dot{z}-L,\label{eqn7}%
\end{equation}
where the momenta are%
\begin{equation}
p_{x}=\frac{\delta L}{\delta\dot{x}}=e^{2u\left(  y\right)  }\dot
{x},\label{eqn8}%
\end{equation}%
\begin{equation}
p_{y}=\frac{\delta L}{\delta\dot{y}}\simeq0,\label{eqn8a}%
\end{equation}%
\begin{equation}
p_{z}=\frac{\delta L}{\delta\dot{z}}=e^{-2v\left(  -y\right)  }\dot
{z}.\label{eqn9}%
\end{equation}
By solving (\ref{eqn8}) and (\ref{eqn9}) for the velocities%
\begin{equation}
\dot{x}=e^{-2u\left(  y\right)  }p_{x}~,\label{eqn9a}%
\end{equation}%
\begin{equation}
\dot{z}=e^{2v\left(  -y\right)  }p_{z}~\label{eqn9b}%
\end{equation}
and substituting them into (\ref{eqn7}), we derive the total Hamiltonian,%
\begin{equation}
H_{T}=p_{y}\dot{y}+\frac{1}{2}p_{x}^{2}e^{-2u\left(  y\right)  }+\frac{1}%
{2}p_{z}^{2}e^{+2v\left(  -y\right)  },\label{eqn9c}%
\end{equation}
where $p_{y}$ is the primary constraint (see (\ref{eqn8a})), the time
development of which leads to the secondary constraint,
\begin{equation}
\left\{  p_{y},H_{T}\right\}  =p_{x}^{2}e^{-2u\left(  y\right)  }u^{\prime
}\left(  y\right)  +p_{z}^{2}e^{+2v\left(  -y\right)  }v^{\prime}\left(
-y\right)  \equiv\phi.\label{eqn9d}%
\end{equation}
Further, the time development of the secondary constraint, $\phi$, is
\begin{equation}
\left\{  \phi,H_{T}\right\}  =\dot{y}\left(  p_{x}^{2}e^{-2u\left(  y\right)
}\left[  u^{\prime\prime}\left(  y\right)  -2\left(  u^{\prime}\left(
y\right)  \right)  ^{2}\right]  -p_{z}^{2}e^{+2v\left(  -y\right)  }\left[
2\left(  v^{\prime}\left(  -y\right)  \right)  ^{2}+v^{\prime\prime}\left(
-y\right)  \right]  \right)  .\label{eqn9e}%
\end{equation}
By using the differential condition for functions (\ref{eqn5})-(\ref{eqn6}),
one obtains,%
\begin{equation}
\left\{  \phi,H_{T}\right\}  =\dot{y}\phi\label{eqn10}%
\end{equation}
from (\ref{eqn9e}). There are no new constraints, and the closure of the Dirac
procedure is reached. The primary and secondary constraints have the following
Poisson Brackets (PBs)%
\begin{equation}
\left\{  p_{y},\phi\right\}  =-\phi;\label{eqn11}%
\end{equation}
therefore, all constraints are first-class.

According to the Dirac conjecture \cite{Diracbook}, the knowledge of all
first-class constraints is sufficient to find the gauge transformations. Using
the Castellani theorem \cite{Castellani} the generator for a system with
primary and secondary first-class constraints\footnote{In the literature, the
term `secondary constraint' is often applied to all non-primary constraints.
We use secondary, tertiary, etc., as the length of a chain of constraints
defines the highest order of the time derivative of the gauge parameter in the
generator \cite{Castellani}.} is%
\begin{equation}
G=\dot{\varepsilon}G_{1}+\varepsilon G_{0}~,\label{eqn12}%
\end{equation}
where%
\[
G_{1}=p_{y}%
\]
and%
\begin{equation}
G_{0}=-\left\{  G_{1},H_{T}\right\}  +\alpha p_{y}=-\phi+\alpha p_{y}%
~.\label{eqn15}%
\end{equation}
The parameter $\alpha$ is derived from the condition%
\begin{equation}
\left\{  G_{0},H_{T}\right\}  =PFC\label{eqn16}%
\end{equation}
where $PFC$ signifies primary first-class constraints. For total Hamiltonian
(\ref{eqn9c}) one obtains%
\begin{equation}
\left\{  G_{0},H_{T}\right\}  =-\dot{y}\phi+\alpha\phi.\label{eqn17}%
\end{equation}
From (\ref{eqn17}) it follows that%
\[
\alpha=\dot{y}.
\]

The explicit form of the generator for the considered model is%
\[
G=\dot{\varepsilon}p_{y}+\varepsilon\left(  -\phi+\dot{y}p_{y}\right)  .
\]
The gauge transformations of all phase-space variables can be found; but we
present only the variables of configurational space, which are:%
\begin{equation}
\delta x=\left\{  G,x\right\}  =\varepsilon2p_{x}e^{-2u\left(  y\right)
}u^{\prime}\left(  y\right)  =\varepsilon2\dot{x}u^{\prime}\left(  y\right)
,\label{eqn20}%
\end{equation}%
\begin{equation}
\delta y=\left\{  G,y\right\}  =-\dot{\varepsilon}-\varepsilon\dot
{y},\label{eqn21}%
\end{equation}%
\begin{equation}
\delta z=\left\{  G,z\right\}  =\varepsilon2p_{z}e^{+2v\left(  -y\right)
}v^{\prime}\left(  -y\right)  =\varepsilon2\dot{z}v^{\prime}\left(  -y\right)
,\label{eqn22}%
\end{equation}
where we have substituted (\ref{eqn8}) and (\ref{eqn9}) into (\ref{eqn20}) and
(\ref{eqn22}) to obtain the transformations in configurational space for
comparison with the results of the Lagrangian approach. These transformations
differ greatly from those presented in Eq. (WLW23) by the authors of
\cite{WangLiWang2009}; and the reason for this difference lies in their use of
an incorrect expression for the generator, Eq. (WLW22), in which a product of
the primary first-class constraints with the highest-order time derivative of
the gauge parameter (it depends on the length of constraint chain
\cite{Castellani}) had not been used. The source of this new construction is
unknown to us, it can only be found in articles written by the authors, not in
monographs on constrained dynamics, nor can it be found in any article known
to us (i.e. see \cite{Kurt,GTbook,HTbook,RRbook})\footnote{A possible
exception is the book \cite{Libook}, which we were unable to find.}. Further,
$G_{0}$ in generator (\ref{eqn12}) is not necessarily a pure, secondary
constraint (it is defined in (\ref{eqn15})-(\ref{eqn16})); only under special
circumstances, when PB (\ref{eqn10}) is zero, is this the case.

If the gauge transformations are obtained in the Hamiltonian approach, then
they must also be the gauge transformations of the corresponding Lagrangian.
This general characteristic for this particular model can be checked by using
Noether's second theorem \cite{Noether,Noether-eng}: if gauge transformations
exist, then there is a corresponding DI -- a linear combination of
ELDs\footnote{Recently in \cite{BanRoy} terms quadratic in ELDs were
introduced into DI, in contradiction to the definition of the linearity of the
DIs. The analog of this linearity at the Hamiltonian level is the absence of
the power of constraints. We shall return to this matter in detail in another
paper.}. And if the gauge transformations are known, then such a DI is not
difficult to restore (i.e. see Schwinger \cite{Schwinger}) from
\begin{equation}
\int\left(  L_{x}\delta x+L_{y}\delta y+L_{z}\delta z\right)  dt=\int%
\varepsilon Idt,\label{eqn25}%
\end{equation}
where $L_{x}=\frac{\delta L}{\delta x}$,\textit{ et cetera} are ELDs, $I$ is a
DI, and $\varepsilon$ is a gauge parameter. Substituting transformations
(\ref{eqn21})-(\ref{eqn22}), and singling out the gauge parameter (using
integration by parts), we obtain%
\begin{equation}
I=2\dot{x}u^{\prime}\left(  y\right)  L_{x}+\dot{L}_{y}-\dot{y}L_{y}+2\dot
{z}v^{\prime}\left(  -y\right)  L_{z}\equiv0.\label{eqn26}%
\end{equation}
Its validity can be confirmed by performing a direct substitution of the ELDs
for Lagrangian (\ref{eqn4}):%
\begin{equation}
L_{y}=u^{\prime}\left(  y\right)  e^{2u\left(  y\right)  }\dot{x}%
^{2}+v^{\prime}\left(  -y\right)  e^{-2v\left(  -y\right)  }\dot{z}%
^{2},\label{eqn27}%
\end{equation}%
\begin{equation}
L_{x}=-e^{2u\left(  y\right)  }\ddot{x}-2u^{\prime}\left(  y\right)  \dot
{y}e^{2u\left(  y\right)  }\dot{x},\label{eqn27a}%
\end{equation}%
\begin{equation}
L_{z}=-e^{-2u\left(  -y\right)  }\ddot{z}-2v^{\prime}\left(  -y\right)
\dot{y}e^{^{-2v\left(  -y\right)  }}\dot{z}\label{eqn29}%
\end{equation}
into DI (\ref{eqn26}). DI (\ref{eqn26}) differs from the false DI of
\cite{WangLiWang2009}, $I_{WLW}$, which is given by Eq. (WLW45),%
\begin{equation}
I_{WLW}=2\dot{x}u^{\prime}\left(  y\right)  L_{x}+\dot{L}_{y}+2\dot
{z}v^{\prime}\left(  -y\right)  L_{z}~,\label{eqn30}%
\end{equation}
$I_{WLW}=0$ is not an identity, and consequently the corresponding
transformations Eq. (WLW45) are not a symmetry of the Lagrangian -- its
variation is neither zero nor a total time derivative. Substitution of
(\ref{eqn27})-(\ref{eqn29}) into (\ref{eqn30}) gives%
\[
I_{WLW}=\dot{x}^{2}\dot{y}e^{2u\left(  y\right)  }\left[  u^{\prime\prime
}\left(  y\right)  -2\left[  u^{\prime}\left(  y\right)  \right]  ^{2}\right]
+\dot{z}^{2}\dot{y}e^{-2v\left(  -y\right)  }\left[  -v^{\prime\prime}\left(
-y\right)  -2\left[  v^{\prime}\left(  -y\right)  \right]  ^{2}\right]  ,
\]
which is not zero identically. Taking into account the differential
relationships (\ref{eqn5})-(\ref{eqn6}), one finds%
\[
I_{WLW}=\dot{y}L_{y}~,
\]
i.e. the true DI is%
\[
I_{WLW}-\dot{y}L_{y}\equiv0,
\]
which is exactly the same as our (\ref{eqn26}).

For completeness, we give the result for the variation of Lagrangian
(\ref{eqn4}) using (\ref{eqn20})-(\ref{eqn22}):%
\[
\delta L=\frac{d}{dt}\left(  e^{2u\left(  y\right)  }u^{\prime}\left(
y\right)  \dot{x}^{2}\varepsilon+e^{-2v\left(  -y\right)  }v^{\prime}\left(
-y\right)  \dot{z}^{2}\varepsilon\right)  ~.
\]
Neither of the two different transformations, Eqs. (WLW23) and (WLW51)
presented in \cite{WangLiWang2009}, are gauge symmetries of the Lagrangian;
but ours show that the results of the Hamiltonian and Lagrangian analyses are
consistent. Therefore, the conclusion about Dirac's conjecture for this model
made by the authors of \cite{WangLiWang2009}: \textquotedblleft Dirac
conjecture loses true to this system\textquotedblright\ is wrong; this
Lagrangian is not a counterexample to the Dirac conjecture.

\subsection{Second example}

Let us consider the second model of Section 3, Eq. (WLW53) of
\cite{WangLiWang2009}:%
\begin{equation}
L=\dot{x}\dot{z}+xz-y\dot{z}~. \label{eqn40}%
\end{equation}
It is obvious that both of the transformations presented by the authors (see
Eqs. (WLW54) and (WLW55)) are not symmetries of Lagrangian (\ref{eqn40});
thus, the conclusion drawn about Dirac's conjecture by the authors is
groundless since it is based on a comparison of two transformations that are
both wrong.

To begin, let us find the Hamiltonian formulation, and try to restore the
gauge transformations. We perform the Legendre transformation,%
\begin{equation}
H_{T}=p_{x}\dot{x}+p_{y}\dot{y}+p_{z}\dot{z}-L,\label{eqn41}%
\end{equation}
where the momenta conjugate to the variables of the Lagrangian are:%
\begin{equation}
p_{y}=\frac{\partial L}{\partial\dot{y}}\simeq0,\label{eqn42}%
\end{equation}%
\begin{equation}
p_{x}=\frac{\partial L}{\partial\dot{x}}=\dot{z},\label{eqn43}%
\end{equation}%
\begin{equation}
p_{z}=\frac{\partial L}{\partial\dot{z}}=\dot{x}-y.\label{eqn44}%
\end{equation}
Two velocities, (\ref{eqn43})-(\ref{eqn44}), can be expressed in terms of the
momenta%
\begin{equation}
\dot{z}=p_{x},\label{eqn45}%
\end{equation}%
\begin{equation}
\dot{x}=p_{z}+y,\label{eqn46}%
\end{equation}
substitution of which into (\ref{eqn41}) gives the total Hamiltonian,
\begin{equation}
H_{T}=p_{y}\dot{y}+p_{z}p_{x}-xz+yp_{x}~.\label{eqn50}%
\end{equation}
The time development of the primary constraint,%
\begin{equation}
\phi_{1}\equiv p_{y}~,\label{eqn50a}%
\end{equation}
is:%
\begin{equation}
\dot{\phi}_{1}=\left\{  p_{y},H_{T}\right\}  =-p_{x}\equiv\phi_{2}%
,\label{eqn51}%
\end{equation}
where $\phi_{2}$ is the secondary constraint. Its time development leads to
the tertiary constraint,%
\begin{equation}
\dot{\phi}_{2}=\left\{  -p_{x},H_{T}\right\}  =-z=\phi_{3},\label{eqn52}%
\end{equation}
on which we have closure of the Dirac procedure by the consistency condition
\begin{equation}
\dot{\phi}_{3}=\left\{  -z,H_{T}\right\}  =-p_{x}=\phi_{2}.\label{eqn53}%
\end{equation}
There are no quarterly constraints for this Hamiltonian.

All constraints ($\phi_{1},\phi_{2},\phi_{3}$) are canonical variables, and
all of the PBs among the constraints are zero; consequently, all of the
constraints are first-class. Castellani's procedure \cite{Castellani} should
lead to gauge transformations that are also gauge transformations of the Lagrangian.

The generator of gauge transformations for a model with tertiary constraints
(compare with (\ref{eqn12})) is%
\[
G=\ddot{\varepsilon}G_{2}+\dot{\varepsilon}G_{1}+\varepsilon G_{0}~,
\]
where%
\[
G_{2}=\phi_{1},
\]%
\[
G_{1}=-\left\{  G_{2},H_{T}\right\}  +\alpha\phi_{1}=-\phi_{2}+\alpha\phi_{1},
\]
and%
\[
G_{0}=-\left\{  G_{1},H_{T}\right\}  +\beta\phi_{1}=\phi_{3}-\alpha\phi
_{2}+\beta\phi_{2}.
\]
From the condition%
\[
\left\{  G_{0},H_{T}\right\}  =PFC
\]
the values of parameters $\alpha$ and $\beta$ follow:%
\[
\phi_{2}-\alpha\phi_{3}+\beta\phi_{2}=0,
\]%
\[
\alpha=0,\text{ \ \ \ }\beta=-1.
\]
The generator in terms of the constraints (all of them are present) is%
\begin{equation}
G=\ddot{\varepsilon}\phi_{1}+\dot{\varepsilon}\left(  -\phi_{2}\right)
+\varepsilon\left(  \phi_{3}-\phi_{1}\right)  ,\label{eqn55}%
\end{equation}
and the gauge transformations are:%
\begin{equation}
\delta x=\left\{  G,x\right\}  =-\dot{\varepsilon},\label{eqn56}%
\end{equation}%
\begin{equation}
\delta y=\left\{  G,x\right\}  =-\ddot{\varepsilon}+\varepsilon,\label{eqn57}%
\end{equation}%
\begin{equation}
\delta z=0.\label{eqn58}%
\end{equation}
It is not difficult to check that this is the symmetry of the Lagrangian by
performing a direct variation of (\ref{eqn40}) using (\ref{eqn56}%
)-(\ref{eqn58}),
\begin{equation}
\delta L=-\frac{d}{dt}\left(  \varepsilon z\right)  .\label{eqn59}%
\end{equation}

One may construct Noether's DI based on the known gauge transformations
(similar to what was done in previous Subsection):%
\begin{equation}
I=-\ddot{L}_{y}+\dot{L}_{x}+L_{y}\equiv0,\label{eqn60}%
\end{equation}
where the ELDs that correspond to this Lagrangian are,
\[
L_{y}=-\dot{z},
\]%
\[
L_{x}=-\ddot{z}+z,
\]
which upon their substitution into (\ref{eqn60}) yields zero.

All constraints appear in generator (\ref{eqn55}), in complete agreement with
Dirac's conjecture, and the gauge transformations obtained in the Hamiltonian
analysis are also the symmetry of the Lagrangian. As in the previous example,
the analysis by the authors \cite{WangLiWang2009} is wrong, and their
conclusion that the \textquotedblleft Dirac conjecture is invalid to this
system\textquotedblright\ is groundless.

\subsection{Third example}

The analysis of the Maxwell theory given in Section 4 of \cite{WangLiWang2009}%
, which supposes to illustrate the validity of Dirac's conjecture (that,
according to the authors, is sometimes correct), is wrong. The transformations
obtained by the authors, Eq. (WLW74):%

\begin{equation}
\delta A_{0}(x)=w_{0}(t),\label{eqn70}%
\end{equation}%
\begin{equation}
\delta A_{i}(x)=\dot{w}_{0}(t)\partial_{i}\delta\left(  x-x^{\prime}\right)
,\label{eqn71}%
\end{equation}
are the most novel, compared with the long known gauge invariance of the
Maxwell theory:%

\begin{equation}
\delta A_{0}(\vec{x},t)=\partial_{0}\phi\left(  \vec{x},t\right)
,\label{eqn72}%
\end{equation}%
\begin{equation}
\delta A_{i}(\vec{x},t)=\partial_{i}\phi\left(  \vec{x},t\right)
~.\label{eqn73}%
\end{equation}

Let us state our first observation: the Maxwell Lagrangian, unlike previous
mechanical (finite dimensional) models, is an example of a field theory, and
the gauge parameters are functions of all of the spacetime coordinates
($\phi\left(  \vec{x},t\right)  $, not just a function of time $w_{0}(t)$).
The transformations provided by the authors of \cite{WangLiWang2009} ,
(\ref{eqn70})-(\ref{eqn71}), are not a gauge symmetry of the Maxwell
Lagrangian. We refer the reader, who is unfamiliar with the gauge invariance
of the Maxwell theory, to any textbook where the Maxwell equations are
mentioned. The many differences between (\ref{eqn70})-(\ref{eqn71}) and
(\ref{eqn72})-(\ref{eqn73}) can be easily observed by anyone, and we will not
list them. The Hamiltonian analysis of the Maxwell Lagrangian for the
first-order formulation can be found in \cite{Kurt}, and in Appendix B of
\cite{Ann}; the restoration of the gauge symmetry for the Yang-Mills theory,
which is more complicated, was given in \cite{Castellani} (see p. 365); and
for the Maxwell theory similar calculations are easy to perform, but to the
best of our knowledge they were not published. In fact, were the results of
\cite{WangLiWang2009} correct, they would not have demonstrated the validity
of Dirac's conjecture, as claimed by the authors of the article
(\textquotedblleft result keeps the validity of Dirac conjecture to free
electromagnetic fields\textquotedblright\ \cite{WangLiWang2009});
transformations (\ref{eqn70})-(\ref{eqn71}) are not a gauge invariance of the
Maxwell action. The most mysterious aspect of the authors' treatment of the
Maxwell theory, however, is the appearance of the same wrong result in both
the Hamiltonian, Eq. (WLW61), and the Lagrangian, Eq. (WLW74),
approaches.\textbf{ }

\section{Some additional counterexamples}

All of the counterexamples to the Dirac conjecture to have appeared in the
literature are based upon the Lagrangians of artificial, unphysical models
that are exclusively mechanical (finite dimensional) and not field-theoretical
models. The debates about the older counterexamples continue, and new
counterexamples emerge (some of which were discussed in the previous Section).
The validity of some counterexamples were questioned, e.g. by Rothe and Rothe
\cite{RRbook,RR2003}.

In covariant theories there are no contradictions to Dirac's conjecture, and
the gauge invariance obtained in the Hamiltonian approach from first-class
constraints (although it depends on field parametrisation) is also a gauge
invariance of the Lagrangian. Among the many examples,\ the Cawley Lagrangian
\cite{Cawley1979} and related modifications \cite{Frenkel1980,Cawley1980} are
perhaps the better known, and we shall discuss them in a separate paper as
they are truly pathological. In this Section we consider the oldest
counterexample of Allcock \cite{Allcock1975}, which is available in two
different parametrisations: the original, and a more recent one constructed by
Henneaux and Teitelboim \cite{HTbook}.

\subsection{Example of Allcock}

The Lagrangian proposed by Allcock is the oldest counterexample
\cite{Allcock1975} (see Eq. (115) on p. 520) and we shall consider it first in
the original choice of variables, i.e.%
\begin{equation}
L=\frac{1}{2}y\dot{x}^{2}. \label{eqn80}%
\end{equation}
We apply the same procedure as that used in the previous Section; performing
the Legendre transformation%
\begin{equation}
H_{T}=\pi_{y}\dot{y}+\pi_{x}\dot{x}-L, \label{eqn81}%
\end{equation}
with the generalised momenta,%
\begin{equation}
\pi_{y}\simeq0, \label{eqn82}%
\end{equation}%
\begin{equation}
\pi_{x}=y\dot{x}. \label{eqn83}%
\end{equation}
Then solving (\ref{eqn83}) for the velocity,%
\begin{equation}
\dot{x}=\frac{\pi_{x}}{y}, \label{eqn84}%
\end{equation}
we obtain the total Hamiltonian:%
\begin{equation}
H_{T}=\pi_{y}\dot{y}+\frac{1}{2}\frac{\pi_{x}^{2}}{y}. \label{eqn85}%
\end{equation}
The consistency condition for the primary constraint, $\pi_{y}$, yields the
secondary constraint,%
\[
\dot{\pi}_{y}=\left\{  \pi_{y},H_{T}\right\}  =\frac{1}{2}\frac{\pi_{x}^{2}%
}{y^{2}}\equiv\chi.
\]
In terms of the primary and secondary constraints, the Hamiltonian
(\ref{eqn85}) is%
\begin{equation}
H_{T}=\pi_{y}\dot{y}+y\chi. \label{eqn90}%
\end{equation}
(There is some similarity to General Relativity, where the entire $H_{T}$ is
proportional to the constraints). The time development of the secondary
constraint,%
\[
\left\{  \chi,H_{T}\right\}  =\frac{\pi_{x}^{2}}{y^{3}}\dot{y}=2\frac{\dot{y}%
}{y}\chi,
\]
gives closure of Dirac's procedure since there is no new, tertiary constraint.

As in the previous Section, we construct the generator:%
\[
G=\dot{\varepsilon}G_{1}+\varepsilon G_{0}~,
\]
where%
\[
G_{1}=\pi_{y},
\]
and%
\[
G_{0}=-\left\{  G_{1},H_{T}\right\}  +\alpha\pi_{y}=-\chi+\alpha\pi_{y}.
\]
The condition%
\[
\left\{  G_{0},H_{T}\right\}  =PFC
\]
leads to%
\[
\left\{  -\chi+\alpha\phi,H_{T}\right\}  =2\frac{\dot{y}}{y}\chi+\alpha\chi
\]
and%
\[
\alpha=-2\frac{\dot{y}}{y}.
\]
The explicit form of the generator for this model is given by%
\begin{equation}
G=\dot{\varepsilon}\pi_{y}+\varepsilon\left(  -\chi-2\frac{\dot{y}}{y}\pi
_{y}\right)  ,\label{eqn99}%
\end{equation}
which yields the following gauge transformations for the variables:
\begin{equation}
\delta y=\left\{  G,y\right\}  =-\dot{\varepsilon}+2\varepsilon\frac{\dot{y}%
}{y},\label{eqn100}%
\end{equation}%
\begin{equation}
\delta x=\left\{  G,x\right\}  =\varepsilon\frac{\pi_{x}}{y^{2}}%
=\varepsilon\frac{\dot{x}}{y}\label{eqn101}%
\end{equation}
((\ref{eqn84}) was taken into account). From transformations (\ref{eqn100}%
)-(\ref{eqn101}) the DI follows,%
\begin{equation}
I\left(  y,x\right)  =\dot{E}_{y}+2\frac{\dot{y}}{y}E_{y}+\frac{\dot{x}}%
{y}E_{x}\equiv0,\label{eqn102}%
\end{equation}
which can be checked by performing a direct substitution of ELDs for
(\ref{eqn80}):
\[
E_{y}=\frac{1}{2}\dot{x}^{2},
\]%
\[
E_{x}=-y\ddot{x}-\dot{y}\dot{x}.
\]
The variation of the Lagrangian using (\ref{eqn100})-(\ref{eqn101}) is a total
time derivative,%
\[
\delta L=\frac{d}{dt}\left(  \frac{1}{2}\dot{x}^{2}\varepsilon\right)  .
\]
The Hamiltonian formulation leads to a gauge symmetry that is also a gauge
symmetry of action (\ref{eqn80}), as required.

For further discussion, the commutator of the two transformations
(\ref{eqn100})-(\ref{eqn101}) is needed, i.e.%
\begin{equation}
\left[  \delta_{1},\delta_{2}\right]  \left(
\begin{array}
[c]{c}%
x\\
y
\end{array}
\right)  =\delta_{\left[  1,2\right]  }\left(
\begin{array}
[c]{c}%
x\\
y
\end{array}
\right)  , \label{eqn110}%
\end{equation}
with the gauge parameter,%
\begin{equation}
\varepsilon_{\left[  1,2\right]  }=-\frac{2}{y}\left(  \varepsilon_{1}%
\dot{\varepsilon}_{2}-\varepsilon_{2}\dot{\varepsilon}_{1}\right)  .
\label{eqn112}%
\end{equation}

Therefore, the oldest counterexample to the Dirac conjecture is incorrect
since both the Hamiltonian (Dirac's approach) and Lagrangian (Noether's DI)
allow one to find the same gauge invariance. We shall consider a more recent
modification of Lagrangian (\ref{eqn80}).

\subsection{The Henneaux-Teitelboim counterexample}

In the book of Henneaux and Teitelboim \cite{HTbook} (Section 1.2.2
\textquotedblleft A Counterexample to the Dirac Conjecture\textquotedblright)
another parametrisation of Lagrangian (\ref{eqn80}) is presented (see Eq.
(1.38) of \cite{HTbook}),%
\begin{equation}
L=\frac{1}{2}e^{z}\dot{x}^{2}, \label{eqnF1}%
\end{equation}
which arises from a change of one variable in (\ref{eqn80}):
\begin{equation}
y=e^{z}. \label{eqnF5}%
\end{equation}
Lagrangians (\ref{eqnF1}) and (\ref{eqn80}) are related by change of variables
(\ref{eqnF5}), which allows one to find a gauge invariance by using the
results of the previous Subsection (this example was also analysed in
\cite{DiStefano1983}, before Castellani's paper was published; but in that
work his ideas were mentioned, and the construction of generators was also
briefly discussed in the spirit of Castellani's results).

The DI for the Allcock counterexample was found (see (\ref{eqn102})), and
because the Henneaux and Teitelboim (HT) counterexample is related to the
Allcock counterexample by a change of variables, we can immediately find the
same DI for the new parametrisation (\ref{eqnF5}). The relation for the ELDs
of the variables affected by such a change is%
\[
E_{z}=E_{y}\frac{\partial y}{\partial z}=E_{y}e^{z},
\]
and the substitution of $E_{y}=E_{z}e^{-z}$ and (\ref{eqnF5}) into
(\ref{eqn102}) gives%
\begin{equation}
I\left(  x,z\right)  =\dot{E}_{z}e^{-z}+\dot{z}E_{z}e^{-z}+\dot{x}e^{-z}%
E_{x}\equiv0, \label{eqnF6}%
\end{equation}
which is \textit{the same DI} (and gauge invariance) written in the new
parametrisation. One can check the correctness of (\ref{eqnF6}) by a direct
substitution of the ELDs for (\ref{eqnF1}).

In the Dirac approach, however, such a symmetry cannot be derived (because of
the parametrisation dependence of Dirac's procedure \cite{KKK-5}); but we can
easily find the anticipated result (without even performing the Hamiltonian
analysis) by converting this DI (\ref{eqnF6}) to a form in which the leading
ELD (with higher order time derivative, see \cite{KKK-5}) is free of a
field-dependent coefficient. By multiplying (\ref{eqnF6}) by $e^{z}$ one
obtains%
\begin{equation}
\tilde{I}\left(  x,z\right)  =\dot{E}_{z}+\dot{z}E_{z}+\dot{x}E_{x}%
\equiv0,\label{eqnF10}%
\end{equation}
which is obviously also a DI, but a \textit{different} one since the
corresponding gauge transformations are:%
\begin{equation}
\delta z=-\dot{\varepsilon}+\dot{z}\varepsilon,\label{eqnF11}%
\end{equation}%
\begin{equation}
\delta x=\dot{x}\varepsilon.\label{eqnF12}%
\end{equation}
These transformations, which can just be read from DI (\ref{eqnF10})
(performing operations (\ref{eqn25})-(\ref{eqn26}) in inverse order) keep the
Lagrangian invariant,%
\[
\delta L=\frac{d}{dt}\left(  \frac{1}{2}\varepsilon e^{z}\dot{x}^{2}\right)  .
\]

These gauge transformations, (\ref{eqnF11})-(\ref{eqnF12}), differ from
(\ref{eqn100})-(\ref{eqn101}) because passing from DI (\ref{eqn102}) to DI
(\ref{eqnF10}) is not just based upon a change of variables; the DI was
modified. In particular, this modification is reflected in the commutator of
the two new transformations:%
\begin{equation}
\left[  \delta_{1},\delta_{2}\right]  \left(
\begin{array}
[c]{c}%
x\\
z
\end{array}
\right)  =\delta_{\left[  1,2\right]  }\left(
\begin{array}
[c]{c}%
x\\
z
\end{array}
\right)  \label{eqnF20}%
\end{equation}
with%
\begin{equation}
\varepsilon_{\left[  1,2\right]  }=-\left(  \varepsilon_{1}\dot{\varepsilon
}_{2}-\varepsilon_{2}\dot{\varepsilon}_{1}\right)  . \label{eqnF22}%
\end{equation}
Note that $\varepsilon_{\left[  1,2\right]  }$ is now field independent,
contrary to (\ref{eqn112}).

The Lagrangian parametrisation (\ref{eqnF1}) was discussed in \cite{RR2003}
and the transformations (\ref{eqnF11})-(\ref{eqnF12}) were given in
\cite{RRbook}, although a different method was used to derive them. We
considered only one modification of DI (\ref{eqnF6}), but many others can be
constructed that lead to different gauge transformations.

To check which symmetry of the Lagrangian is produced by the Dirac approach,
we perform the Hamiltonian analysis of (\ref{eqnF1}). Let us find the
constraints, use the Castellani procedure to derive the gauge transformations,
and then compare them with (\ref{eqnF11})-(\ref{eqnF12}).

As before, performing the Legendre transformation of (\ref{eqnF1}),%
\[
H_{T}=\dot{x}p_{x}+\dot{z}p_{z}-L,
\]
where the conjugate momenta,%
\[
p_{z}\simeq0,
\]%
\[
p_{x}=e^{z}\dot{x},~\dot{x}=e^{-z}p_{x},
\]
leads to the total Hamiltonian%
\[
H_{T}=\dot{z}p_{z}+\frac{1}{2}e^{-z}p_{x}^{2}~.
\]
The time development of the primary constraint yields the secondary
constraint,%
\[
\left\{  p_{z},H_{T}\right\}  =\frac{1}{2}e^{-z}p_{x}^{2}\equiv\chi\simeq0,
\]
the time development of which gives closure of the Dirac procedure,%
\begin{equation}
\left\{  \chi,H_{T}\right\}  =-\frac{1}{2}e^{-z}p_{x}^{2}\dot{z}=-\dot{z}%
\chi.\label{eqnF30}%
\end{equation}
Note that all constraints are first-class because of (\ref{eqnF30}) and%
\[
\left\{  p_{z},\chi\right\}  =\chi.
\]
Using the Castellani procedure, we find the gauge generator,
\[
G=\dot{\varepsilon}G_{1}+\varepsilon G_{0}~,
\]
where%
\[
G_{1}=p_{z},
\]%
\[
G_{0}=-\left\{  G_{1},H_{T}\right\}  +\alpha p_{z},
\]
with $\alpha$ defined from the condition,%
\[
\left\{  G_{0},H_{T}\right\}  =PFC,
\]
which gives%
\[
\left\{  G_{0},H_{T}\right\}  =\dot{z}\chi+\alpha\chi
\]
and $\alpha=-\dot{z}$. Therefore, the explicit form of the generator is%
\begin{equation}
G=\dot{\varepsilon}p_{z}-\varepsilon\left(  \frac{1}{2}e^{-z}p_{x}^{2}+\dot
{z}p_{z}\right)  ,\label{eqnF40}%
\end{equation}
and the transformations of the variables are:%
\[
\delta x=\varepsilon e^{-z}p_{x}=\varepsilon\dot{x},
\]%
\[
\delta z=-\dot{\varepsilon}+\varepsilon\dot{z},
\]
which are the same as those in (\ref{eqnF11})-(\ref{eqnF12}) (found by a
modification of the DI) .

\subsection{The Natural Parametrisation of the Allcock Lagrangian}

We have shown that although the Allcock and HT examples are considered
distinct counterexamples to the Dirac conjecture, they are in fact related by
a change of field parametrisation. For the Allcock Lagrangian (\ref{eqn80})
the commutator for two transformations leads to a field-dependent gauge
parameter (\ref{eqn112}); while for the HT Lagrangian, gauge parameter
(\ref{eqnF22}) is field independent. At this point, we wish to find the
\textquotedblleft natural\textquotedblright\ parametrisation of the Allcock
Lagrangian by a general method (see \cite{KKK-5}). The natural parametrisation
is a choice of variables for this Lagrangian, which leads to gauge
transformations with the simplest commutator, and we want to explore the
possibility of finding such a choice of variables.

Let us return to the Allcock Lagrangian (\ref{eqn80}) and modify DI
(\ref{eqn102}), i.e. to find another gauge transformation for this Lagrangian.
Multiplying (\ref{eqn102}) by $f(y)$, a function of $y$, we obtain
\begin{equation}
\tilde{I}\left(  x,y\right)  =f\left(  y\right)  \dot{E}_{y}+2f\left(
y\right)  \frac{\dot{y}}{y}E_{y}+f\left(  y\right)  \frac{\dot{x}}{y}%
E_{x}\equiv0.\label{eqnF7}%
\end{equation}
This new DI describes new gauge transformations which are%
\begin{equation}
\delta x=\varepsilon f\left(  y\right)  \frac{\dot{x}}{y},\label{eqnN1}%
\end{equation}%
\begin{equation}
\delta y=-\partial_{0}\left(  \varepsilon f\left(  y\right)  \right)
+2\varepsilon f\left(  y\right)  \frac{\dot{y}}{y}.\label{eqnN2}%
\end{equation}

Let us try to find a condition on function $f(y)$ that will lead to a zero
commutator of two consecutive transformations. The commutator for variable $x$
seems to be simpler to calculate, and it gives rise to%
\begin{equation}
\left[  \delta_{1},\delta_{2}\right]  x=\dot{x}\left(  \frac{f^{\prime}\left(
y\right)  }{y}-2\frac{f\left(  y\right)  }{y^{2}}\right)  f\left(  y\right)
\left(  \varepsilon_{1}\dot{\varepsilon}_{2}-\varepsilon_{2}\dot{\varepsilon
}_{1}\right)  ,\label{eqnN3}%
\end{equation}
which is zero if the expression in brackets is zero, that corresponds to the
ordinary differential equation (ODE),%
\[
\frac{f^{\prime}\left(  y\right)  }{y}=2\frac{f\left(  y\right)  }{y^{2}}%
\]
with the solution%
\begin{equation}
f\left(  y\right)  =ky^{2}\label{eqnN5}%
\end{equation}
($k$ is a constant). For this function $f\left(  y\right)  $, DI (\ref{eqnF7})
becomes%
\begin{equation}
\tilde{I}\left(  x,y\right)  =ky^{2}\dot{E}_{y}+2ky\dot{y}E_{y}+ky\dot{x}%
E_{x}\equiv0.\label{eqnF7a}%
\end{equation}

In the previous Subsection, we modified the DI to find the symmetries that the
Hamiltonian produces for the new parametrisation. Here the situation is
different; we want to keep this DI, as it describes the simplest
transformations with a zero commutator (we were looking for such a
possibility), but change the parametrisation in such a way that this symmetry
will follow from the Hamiltonian analysis (and therefore provide a natural
parametrisation). We seek a new parametrisation for which the Noether DI
(\ref{eqnF7a}) will contain a leading ELD with a field-independent
coefficient. Because the leading term in (\ref{eqnF7a}) depends only on one
variable, this is the only variable we have to change, i.e. $y=\tilde
{y}\left(  y\right)  $.

With this change of variable, we have a relation between new and old ELDs,
i.e.%
\begin{equation}
E_{y}=E_{\tilde{y}}\frac{d\tilde{y}}{dy}, \label{eqnN8}%
\end{equation}
and its substitution into (\ref{eqnF7a}) (only first term has to be
considered) yields%
\[
ky^{2}\dot{E}_{y}=ky^{2}\frac{d\tilde{y}}{dy}\dot{E}_{\tilde{y}}+ky^{2}%
\frac{d}{dt}\frac{d\tilde{y}}{dy}E_{\tilde{y}}~,
\]
where only the first term, with a derivative of the ELD, is of concern. The
coefficient is equal to one if%
\[
ky^{2}\frac{d\tilde{y}}{dy}=1,
\]
which is again an ODE with the solution,%
\begin{equation}
\tilde{y}=-\frac{1}{ky}. \label{eqnN10}%
\end{equation}

Function $\tilde{y}$ provides a parametrisation for which the Hamiltonian
formulation will produce gauge transformations with a zero commutator. Before
considering the Hamiltonian formulation, we substitute (\ref{eqnN10}) into
(\ref{eqn80}) (for simplicity, let us take $k=-1$) to obtain%
\begin{equation}
L=\frac{1}{2\tilde{y}}\dot{x}^{2},\label{eqnN12}%
\end{equation}
and using (\ref{eqnN8}) and (\ref{eqnN10}), find DI (\ref{eqnF7a}) for this
parametrisation:%
\begin{equation}
\tilde{I}\left(  x,\tilde{y}\right)  =\dot{E}_{\tilde{y}}-\frac{\dot{x}%
}{\tilde{y}}E_{x}\equiv0,\label{eqnN14}%
\end{equation}
which can be checked by substituting the ELD for Lagrangian (\ref{eqnN12}).
The gauge transformations follow:%
\begin{equation}
\delta x=-\frac{\dot{x}}{\tilde{y}}\varepsilon,\label{eqnN16}%
\end{equation}%
\begin{equation}
\delta\tilde{y}=-\dot{\varepsilon}.\label{eqnN17}%
\end{equation}
It is easy to check that the commutator for (\ref{eqnN16})-(\ref{eqnN17}),
unlike the commutator for the previous parametrisations (see (\ref{eqn80}) and
(\ref{eqnF1})), is zero:%
\[
\left[  \delta_{1},\delta_{2}\right]  \left(
\begin{array}
[c]{c}%
x\\
\tilde{y}%
\end{array}
\right)  =0,
\]
and the same is true for (\ref{eqnN5}) (the DI remains the same).

We shall confirm the above results by briefly considering the Hamiltonian
formulation of (\ref{eqnN12}). The total Hamiltonian is%
\begin{equation}
H_{T}=\overset{\cdot}{\tilde{y}}p_{\tilde{y}}+\frac{1}{2}\tilde{y}p_{x}%
^{2}~,\nonumber
\end{equation}
where $p_{\tilde{y}}$ is the primary constraint ($\dot{x}=\tilde{y}p_{x}$),
the time development of which leads to the secondary constraint:%
\begin{equation}
\left\{  p_{\tilde{y}},H_{T}\right\}  =-\frac{1}{2}p_{x}^{2}\equiv
\chi.\nonumber
\end{equation}
The Dirac closure is simple%
\[
\left\{  \chi,H_{T}\right\}  =0.
\]
Further, the PBs among the constraints are zero, and the total Hamiltonian has
the simple form,%
\[
H_{T}=\overset{\cdot}{\tilde{y}}p_{\tilde{y}}-\tilde{y}\chi.
\]

The Castellani procedure for building the generator is also considerably
simplified,%
\[
G=\dot{\varepsilon}p_{\tilde{y}}+\varepsilon G_{0}~,
\]
with%
\[
G_{0}=-\left\{  \phi,H_{T}\right\}  +\alpha\phi=-\chi+\alpha\phi.
\]
The condition%
\[
\left\{  G_{0},H_{T}\right\}  =PFC
\]
gives%
\[
\alpha=0;
\]
therefore, the generator is%
\begin{equation}
G=\dot{\varepsilon}p_{\tilde{y}}+\varepsilon\chi,\label{eqnN30}%
\end{equation}
which leads to transformations (\ref{eqnN16})-(\ref{eqnN17}).

We have shown that it is possible to make a change of parametrisation for the
Allcock Lagrangian to convert it to a natural form for which the commutator of
the gauge transformations is the simplest possible, i.e. equal to zero; the
algebra of constraints is also the simplest (all PBs are zero); and as
consequence of this simple algebra of constraints, the Castellani procedure
leads to the simplest generator (\ref{eqnN30}), where $G_{0}$ is equal to the
secondary constraint, while the more complicated expressions for $G_{0}$
appear in (\ref{eqn99}) and (\ref{eqnF40}).

\section{Conclusion}

The construction of counterexamples to the Dirac conjecture seems to have been
a never ending theme for over three decades. Many specially devised examples
are not counterexamples at all, they are just a result of wrong analysis, as
in the paper by Wang, Li, and Wang \cite{WangLiWang2009}. Three wrong
examples, which had been presented by these authors, were considered in
Section II, and the sources of their numerous mistakes described. Correct
analysis shows that a consistent result can be obtained: the generator, built
from all first-class constraints, produces a gauge symmetry that is also a
gauge symmetry of a Lagrangian.

The oldest example, due to Allcock \cite{Allcock1975}, was considered in
Section III, and it also leads to a consistent result. Because this example is
presented in two parametrisations it allows an explicit demonstration of the
parametrisation dependence of the Dirac method, something we have already
discussed for some mechanical models \cite{KKK-5} and for field theory
(different field parametrisations of General Relativity) \cite{KKK-1,KKK-3}.
We have also demonstrated how the natural parametrisation for the Allcock
model can be constructed by using, as a criterion, the choice of the simplest
properties of the commutator of two transformations.

According to \cite{Pathol}: \textquotedblleft The term `pathological'\ is used
in mathematics to refer to an example specifically cooked up to violate
certain almost \textit{universally valid} properties\textquotedblright%
\ (italicisation is ours).\ If a particular parametrisation was
\textquotedblleft cooked up\textquotedblright\ for a theory for which the
Dirac conjecture is valid,\ it should be no surprise that problems popped up.
For example, if a non-covariant change of variables is performed in a
generally covariant theory (e.g. Einstein-Hilbert (EH) action), could one
reasonably expect covariant results? Is such a result grounds for one to
conclude that EH is \textit{not} a generally covariant theory? The same logic
applies to the Dirac conjecture: it is valid for the natural parametrisation
of a model, but it is invalid for some \textquotedblleft cooked
up\textquotedblright\ parametrisations. For example, the Hamiltonian
formulation of EH in natural, metric variables leads to the gauge invariance
of full diffeomorphism \cite{PLA}, but in ADM parametrisation it produces a
different symmetry \cite{Myths}; one set of transformations forms a group and
the other does not \cite{KKK-3}. The same behaviour was observed for the model
of Allcock -- the property of the commutator of two consecutive
transformations is parametrisation dependent, and one can find the
parametrisation that has the simplest, zero-valued commutator.

For some parametrisations the transformations can be found according to the
Dirac conjecture; although it is not necessarily the natural choice (these
transformations might not form a group or the commutator might not be of the
simplest form). But for some parametrisations (the term `pathological
parametrisation' is suitable) the problems are much more severe. For example,
the Dirac procedure does not have unique closure, then\ the Castellani
procedure cannot produce unique transformations (the best known model of this
kind is due to Cawley, which will be discussed elsewhere).

For those models known to us, which have pathological parametrisations,
changes of variables can always be found to return the system to normality;
thus making the Dirac conjecture \textquotedblleft\textit{universally
valid}\textquotedblright. If for some model a change of parametrisation does
not exist for which the first-class constraints can be unambiguously found
(after phase-space reduction, i.e. solving the second-class constraints), then
the Dirac conjecture simply cannot be applied, and this should be considered
as a strong indication that such models could be ill-defined.

\section{Acknowledgment}

We would like to thank A. M. Frolov, D. G. C. McKeon, and A.V. Zvelindovsky
for discussions and M. Nowak for attracting our attention to paper
\cite{WangLiWang2009}.

\end{document}